# Perceived irrelevance and mastery vs. performance achievement goals: Two mindset variables within attitudinal experiences of life science majors in introductory physics

Andrew J. Mason

*Department of Physics and Astronomy, University of Central Arkansas,*
*201 S. Donaghey Avenue, Conway, AR 72035*

**Abstract:** In a previous study, students' self-expressed learning orientations towards an exercise centered on self-monitoring one's ability to solve a pre-lab physics problem were identified from a post-test feedback survey given to an introductory algebra-based physics student population spanning six measured semesters, and examined as a potential variable in course performance, force and motion conceptual understanding, and attitudes towards learning physics. The sampled population, which primarily consists of life science majors, was also asked in the same feedback survey to discuss what portion or portions of the course were relevant to their respective choices of major. In this study, we examine the fact that about 50 students out of 218 sampled students, or 23% of the sample population, explicitly stated that they perceived no relevance at all of the course to their respective majors, whereas the other 168 students cited portions of the course or the entirety of the course as being relevant to their majors. A follow-up investigation of perceived relevance versus irrelevance shows that attitudes towards physics will experience more expert-like shifts for students who perceive relevance than students who do not; in particular, the attitudinal survey's item clusters that pertain to personal interest and real-world connections appear to show the strongest effect. Further examination showed that biology majors and health science majors (two distinctive sub-populations of life science majors) show similar pre-post trends for relevance vs. irrelevance perceptions, whereas students with a performance achievement goal appeared to bifurcate between a novice-like shift for perceived irrelevance and no attitudinal shift from pre to post for perceived relevance. Discussion includes emphasis on limitations imposed by institution type among other factors.

## I. INTRODUCTION

The notion of student mindset, [1-2] as recently applied to learning physics, [3-4] builds upon research upon students' epistemological views of learning, [5] particularly with regard to learning physics. [6] In particular, the notion of student mindset has been recognized as a potential factor for the success of physics majors, both throughout the undergraduate physics major curriculum [7] and extending into graduate school and even career paths for physical science majors. [8] The general consensus of the literature appears to be that seeing oneself as a physicist does have an influence upon being successful as a physicist.

### A. Two mindset variables: achievement goals and perceived relevance

The topic of student mindset, primarily with regard to student beliefs about intelligence, has a very strong presence in psychology literature. [1,9] We focus on two aspects of student mindset in this paper. First, in previous literature, [10-13] we discussed the effect of achievement goals, specifically mastering vs. performance achievement goals, on an introductory algebra-based physics course. In this paper, we consider achievement goals alongside a second variable, namely perceived relevance of physics to students' respective majors. [14] Perceived relevance has been duly noted in terms of students' buy-in and self-perception, [15] and clearly identified as a potential issue when interpreting students' content gains. While many other factors of mindset are important to consider for introductory physics, [4] this paper will focus on the combined effect of mastery/performance achievement goals and perceived relevance of physics to one's life science major.

The above concerns come to the forefront when considering a research-based implementation of curricular change. Hypothetically, students will become more engaged and have better performance results due to the implementation (as compared to a traditional course plan), but if student mindset variables are not optimal, then the course population may not benefit from the research-based implementations. As an example from previous work, [10] students in an introductory algebra-based first-semester physics course were presented with a metacognitive coordinated group problem solving task [16] working on a context-rich problem [17] at the beginning of their laboratory sections, which was followed by a metacognitive "self-diagnosis" exercise for individual students to consider their respective strengths and weaknesses at the problem solving exercise [18-19]. In a post-test feedback survey, students self-reported how the exercise helped them in the course; most of them expressed learning orientations that were either mastery-based or performance-based achievement goals [9]. A clear difference between these two achievement goals was visible in results on a validated survey regarding attitudes towards physics. [20]

While the average course GPA across both learning orientation groups in the Mason and Bertram study was virtually identical, a borderline result showed that students who had either achievement goal performed marginally better in the course than students who did not express an achievement goal. This result appears to reflect earlier, more detailed studies that also show that, within the scope of a single introductory course, it is simply more important to have either achievement goal than to have neither. For example, findings within a larger sample of introductory psychology students showed that students with primarily mastery orientation, students with primarily performance orientation, and students who were highly motivated in both mastery and performance, all had similar outcomes on achievement measurables, while low-motivated students (akin to the "vague" orientation in Ref. 10) had significantly poorer results on achievement. [21]

While seemingly modest, the Ref. 10 results suggest that students' respective mindsets may either assist or undermine the pedagogical benefit they receive from research-based instruction. In addition, the algebra-based course was designed for non-physical-science STEM majors, predominately biology and health science majors. It was determined that health science majors tended to fare more poorly (e.g. on FCI and course performance), as well as have more novice-like attitudes on the CLASS (both pre-test and post-test), than did biology majors.

Between the variable of type of life science major and the variable of achievement goals, it becomes clear that a) certain subpopulations of the student sample may not receive the intended benefit of research-based interventions if they are not explicitly considering mastery of the subject as a goal; and b) students that are more removed from the natural science setting of introductory physics (i.e. the health science majors, whose concentrations ranged from health administration to pre-physical/occupational therapy to nutrition, etc.) will on average have a harder time being prepared for the course from start to finish than will a life science major that is more clearly linked to natural science (i.e. the biology majors, some of whom were pre-medical and some of whom were not).

For this paper, we revisit the same population described within the Mason and Bertram study in order to examine perceived relevance of physics to students' chosen major within this population. In combination with achievement goals, perceived relevance may also help to predict whether students' perceptions of learning physics will be more novice-like or expert-like by the end of a first-semester introductory course's instruction. It therefore becomes important both to identify perceived relevance within a population of life science majors, and to examine the variable within the other variable of self-expressed achievement goal (mastery, performance, or neither).

### B. Theoretical framework for implementation

The implementation of the problem-solving framework within both this study and the Mason and Betram study rests upon the notion of cognitive apprenticeship, [22] specifically the modeling-coaching-fading model for mentoring a student into mastery of the topic. Scaffolding of some kind is necessary to ease the students' learning progress as well, e.g. in terms of balancing innovation with efficiency. [23] In this case, a rubric which was designed to metacognitively "self-diagnose" errors on a quiz [18-19] was adapted to allow individual students to identify their own strengths and weaknesses for a given problem attempt. The need to sustain the intervention from week to week, in order to reinforce students' practices, is also apparent, as a one-time intervention may not be enough to reinforce the exercise's metacognitive skills [18-19]. Therefore, the metacognitive exercise may have better results if applied on a weekly basis, e.g. in a recitation setting for a weekly problem solution attempt, and adapted accordingly from a "self-diagnosis" quiz setting to a self-monitoring task in a worked example setting.

In addition, coordinated group problem solving [16] offers scaffolding in the form of groups working together in a proactive setting. Measures must be taken to prevent the typical pitfall of group work, namely that one student ends up dominating the workload while others do not contribute (and subsequently, do not learn the material as well). The model for this task, in keeping with the "Minnesota model" for problem solving, [24] suggests that students be given "roles" to use and develop these roles mutually with their partners. Alternatively, one may also provide incentive for students to work individually while being allowed to consult with their partners as needed.

Before examining these measures for performance, it must also be considered that students' focus on learning may be affected by certain circumstances, e.g. majoring in something other than physical science (see the next section). Aspects of student mindset, such as achievement goals and perceived relevance, become potential factors in students' overall performance. Perceived relevance has recently become a focus within PER literature. More recent efforts include a focus on perceived relevance within well-known attitudinal surveys in physics [25] or in terms of pre-service physics teachers' issues with their own perceived relevance of physics to their teaching careers. [26] In particular, it is a concern that life science majors, e.g. pre-medical students, may see a decline in perceived relevance over the course of a semester, as explicitly detailed by individual items on an attitudinal survey for physics. [27-28]

A feedback survey that investigates students' perceived relevance directly, as well as students' achievement goals indirectly, was used to collect data for achievement goals in the previous study. [10] We follow up in this paper on the more direct single-question measurement of students' perceived relevance of physics to their respective majors.

### C. Considerations and Limitations for Student Mindset Research at the Host Institution

There are several considerations that must be understood for the context of the following paper. First, not all students in introductory physics are physics majors, or for that matter physical science and engineering majors, such that it is not immediately apparent to students outside of these majors why they need to be in an introductory physics course. A strong example of research on this issue exists within the Introductory Physics for Life Science (IPLS) literature in recent years, [29-31] e.g. in the concerns for pre-medical biology majors and pre-physical therapy health science majors. There is extensive literature regarding addressing student mindset – e.g. interviews of biology majors to check for perceived relevance of physics to their career paths [32], as well as coordination with biology faculty for input on which physics topics would more ideally assist biology majors at the introductory level [33]. The notion that performance in introductory physics may be influenced by students' attitudes and interest in physics has been well established in recent literature, e.g. in the setting of a private liberal arts college [34-35] where an IPLS transformation has been shown to improve students' attitudes towards physics.

Second, there is more generally an issue within the imbalance of literature on introductory physics courses, as discussed by Kanim and Cid. [36] For example, among the results found in this study were an overemphasis on research for introductory physics courses concentrated within calculus-based physics, with relatively little focus on algebra-based physics (as well as a very small focus on community college and high school sections of physics). However, many introductory physics courses constructed with a primarily life-science major population are constrained to algebra-based physics contexts, whether administratively or by departmental decision, and nevertheless must teach scientific skills and content knowledge whether calculus is explicitly invoked or not.

Third, the vast majority of published IPLS research comes from researchers at either large or very large research institutions, with the remaining researchers tending to be based within private liberal arts colleges or similar institutions. The above institution types tend to have very high tuition rates, high standards for acceptance, and relatively low acceptance rates. Other institution types, e.g. regional state universities with higher acceptance rates and lower tuition costs, have relatively little literature in the field. A potential pitfall of this situation is that the solutions proposed by the literature for large research universities do not necessarily work for students who attend other institution types, e.g. because the institutional context of the literature does not match the institutional context of other institutions.

The latter two points are of importance due to their influence on the study, namely that the host institution is a primarily-undergraduate state university with regional focus, and the host department of physics and astronomy is undergraduate-only. This means that certain factors within much of the literature cannot be directly replicated. One factor is that simply not having a graduate program means that graduate teaching assistants are not available; undergraduate learning systems are primarily only available for tutoring hours. The host department otherwise does not offer recitation sections and must have faculty teach all laboratory sections. On the other hand, this means that individual faculty members can have full control over in-class interventions, albeit within the limitations of traditional lecture-lab instruction models for large introductory courses.

A second factor is that calculus-based introductory physics is restricted to courses taken by physics majors and a few other majors closely associated with physics (e.g. ACS-accredited tracks within the chemistry major). The life science majors must instead take the introductory algebra-based physics course that is reserved for all other STEM majors. Many IPLS studies at large research universities are largely done on calculus-based IPLS courses, but this is often not possible at other institution types.

As a third factor, not all life science majors at the host institution are the same specific major, as is the case in much of the IPLS literature. Life science majors are split between majors within biology tracks and health science majors within a separate college. Biology majors include both pre-professional concentrations, e.g. pre-medical or pre-dental, and regular biology majors without a pre-professional track. Majors within the health science college includes pre-physical therapy and pre-occupational therapy tracks, as well as several different majors outside of these tracks, e.g. nutrition and health administration. Biology majors and health science majors are roughly equal in population size for most given algebra-based introduction sections. The logical solution is to make two separate course sections that cater respectively to these two sets of majors; however, before this can occur, student mindset issues must be addressed for each set of majors in order to optimize the experience in both course sections.

## II. CURRENT RESEARCH GOALS

In this study, which is a follow-up to Mason and Bertram, [10] we examine a question given in a post-test feedback survey: "Which sections of the [course] material did you find related well to the coursework in your major?" While students cited many items within the course that they perceived to be relevant to their respective majors, 50 out of 218 total students stated that they saw no relevance whatsoever to their perceived majors. We therefore look at the major variable of "relevance" versus "irrelevance", i.e. students perceiving any kind of relevance at all vs. students who perceived no relevance.

In this paper, we will attempt to answer the following research questions. First, we will see if perceived relevance has a quantitative effect on FCI pre-post conceptual understanding. Second, we will determine if perceived relevance has an effect on pre-post attitudinal shifts measured by the CLASS survey, both overall and on seemingly pertinent item clusters (e.g. Personal Interest and Real-World Context). Third, we will look within-groups for both choice of life science major (either biology or health science) and achievement goal orientation (Framework, Performance, or Vague), to see if any of these subgroups is particularly affected by perceived relevance.

## III. PROCEDURE

### A. Classroom Setup and Instruments

From the period of Spring 2014 to Spring 2017, six semester sections from a single instructor's first-semester introductory algebra-based physics course were sampled. The instructor taught within the context of a traditional lecture-lab format with no recitations, directly teaching all laboratory sections (typically two or three sections per semester, 24 students each) as well as a single lecture section attached to all laboratory sections (48-72 students total). Instructor bias between sections is therefore not a factor.

The instructor took care to introduce life science applications of each topic covered in the first-semester section (kinematics, linear dynamics, impulse and momentum, work and energy, rotational statics, rotational kinematics and dynamics, limited thermodynamics topics, and static fluids). This life science coverage would consist of a percentage of in-class worked examples and homework problems.

Since there were no recitation sections, the instructor used the first 50-60 minutes to implement a pre-lab problem exercise (see Fig.s 1 and 2; also see Ref. 10). The exercise consisted of presenting students with a context-rich problem [17] that was conceptually connected to the lab experiment that would follow, and having them work on the problem for approximately 45 minutes. The students would ideally work in coordinated groups; [16] however, the situation often reflected individual efforts with some students helping out others as needed. [12] During this time, the instructor would be available for questions, clarifications, etc. while making sure all students in each group were still working towards a solution.

At the end of the 45-minute session, the instructor would go over the solution using Socratic dialogue with students (who could then use their worked problem solution attempts to answer). Finally, students would spend a few minutes filling out a self-diagnosis rubric, in which they identified strengths and weaknesses to their problem solving framework attempts. [18-19] Once students individually discussed their rubrics with the instructor, they could proceed with the featured lab experiment with their lab partners. Course points were awarded for active participation rather than completion, so as to avoid making the exercise grade-dependent.

### B. Data Collection

The quantitative data consisted of pre-post student responses on the Force Concept Inventory (FCI) [37] and Colorado Learning Attitudes about Science Survey (CLASS). [20] These surveys were respectively given during the first 50 minutes of each lab section during the first and last weeks of the semester, for pretest and posttests respectively.

Also given as a posttest was a short-answer feedback survey that asked students several questions about their experience in the course. There were four main questions. The first two questions [10] were used to classify students into their expressed mastery vs. performance achievement goals. The first question was "In what ways did you find this [problem-solving] exercise useful towards learning the material in the course?", to which students would respond in terms of how they were mastering a problem-solving framework (or "Framework orientation"), in terms of how they believed it helped them perform better on an exam, homework, or other aspect of the course ("Performance orientation"), or in terms of something else besides a learning goal ("Vague orientation"). Two raters sorted the students' answers into one of these three orientation categories; if they were unsure about a particular student, then the student's answer to the second question, "Do you have any suggestions to make this exercise more useful toward learning the material in the course?" typically served as a means of clarifying the student's primary orientation when answering both questions. The raters were able to come to at least 90% agreement on a randomly-selected subsample of the student population.

The third question of the survey, which will be more of the focus of this paper, was first ask the students what their major was, and then to ask, "Which sections of the material did you find relate well to the coursework in your major?" Students' answers typically focused on a single section of the course material (e.g. forces or torque), on skills pertinent to their majors (e.g. problem solving skills), or on topics in their majors to which they believed the course applied (e.g. the human body). Other responses included "everything in the course" or alternately "nothing in the course." This last response will be the focus on this paper, by which we will define whether students perceived any relevance to their course (for which any other response would be valid) or no relevance at all (for which they would specifically state that they perceived no relevance).

| You are studying a herd of bighorn sheep in Alberta, and have found two rams butting heads with each other to establish dominance. Your camera is able to get a high-speed recording of the rams as they take a running start toward each other, on level ground and along a straight line path, before they collide head-on. This way you can use video analysis software to determine how fast each ram was going before the collision. The first ram looks to be about average size for an adult, while the second ram looks to be about ½ the size of the first ram, so you estimate the mass of the first ram to be about 90 kg* and the second ram to be about 45 kg. In your analysis, you see that the second ram was running much faster (2.10 m/s) than the first ram was (0.90 m/s), such that when they butted heads, the smaller ram was brought to a halt. As a result, you find that you can determine the velocity of the larger ram after the collision. |
|---|

**FIG. 1**. Sample context-rich problem used for student meta-cognition during the first portion of weekly laboratory sections.

| Circle how much you think you understood on: | What were your strengths on each of these parts? | What did you struggle with on each part? |
|---|---|---|
| **Problem description** <br><br> Full/Partial/None | | |
| **Solution construction** <br><br> Full/Partial/None | | |
| **Logical progression** <br><br> Full/Partial/None | | |

**FIG. 2**. Adaptation of the rubric used in Yerushalmi et al. (2012a-b) for the individual metacognition portion of the problem solving exercise.

A total of 218 students over the six sampled semesters submitted a complete data set of the above items; students who did not (e.g. missing class for the pretests) were omitted. Other students were also omitted if they failed the course outright, or if at least one of their survey responses clearly demonstrated that they were not taking a particular survey seriously (e.g. the CLASS survey's item that requests students to answer a particular choice in order to show they are paying attention to the items).

## IV. RESULTS

### A. Distribution of Perceptions of Relevance vs. Irrelevance

The distributions of the student sample into "relevant" (i.e. students who mentioned that they found at least some aspect of the course to be relevant to their majors) and "irrelevant" (i.e. students who stated that they found nothing to be relevant) groups, as well as into subgroups by choice of major and by stated learning orientation, are in Table 1. The learning orientations, as discussed in Ref. 10, are derived from students' answers to how they found the in-class pre-lab problem solving exercise useful; students' responses could be interpreted in a mastery-vs.-framework achievement goal set: mastery of the problem-solving framework presented to them (i.e. Framework), performance on various in-class measurements e.g. exams (i.e. Performance), or not really expressing any learning goal (i.e. Vague).

TABLE 1. Distributions of students who saw the physics course as relevant in some way to their majors vs. students who saw the course as irrelevant. Displayed both in terms of raw number and in terms of the percentage out of each group.

| Group (n) | Relevant | Irrelevant | Irr. % |
|---|---|---|---|
| All (218) | 168 | 50 | 22.9% |
| Biology (91) | 76 | 15 | 16.5% |
| Health (85) | 66 | 19 | 22.4% |
| Other Sci (35) | 21 | 14 | 40.0% |
| Non-Sci (7) | 5 | 2 | 28.6% |
| Framework (76) | 59 | 17 | 22.4% |
| Performance (79) | 63 | 16 | 20.3% |
| Vague (63) | 46 | 17 | 27.0% |

While the percentage of students who perceive irrelevance does not statistically change between groups for choice of major or for achievement goal, slight differences suggest trends that can be checked on validated survey instruments. For example, it appears that a slightly higher percentage of Vague-oriented students (who expressed no achievement goal) perceive irrelevance than do either of the groups who did express either achievement goal. This mirrors the borderline significance finding in Mason and Bertram [10] that students with a Vague orientation had a slightly lower course GPA than did the Framework and Performance orientations. Also, since the percentage of biology majors perceiving relevance is slightly higher than the percentage of health science majors perceiving relevance, there is some sort of effect as well. Having said all this, it is more appropriate to determine which items on a validated attitudinal survey such as the CLASS survey will exhibit the strongest effect with this variable. We also wish to see whether it corresponds to conceptual understanding of force and motion with the FCI survey.

### B. Effect of Relevance on FCI Performance

We next examine whether perceived relevance is a variable on conceptual understanding of force and motion. Table 2 examines the FCI pre-test, post-test, and normalized gain average for the "Relevant" population versus the "Irrelevant" population. The table includes p-values for t-test results between groups; all comparisons were shown to have statistically similar variances (F-test: $p > 0.05$).

Focusing within choice of life science major, an examination of relevance vs. irrelevance did not show any statistical differences in force concept understanding between biology majors and life science majors. This is also the case for looking between achievement goal groups, although a moderate effect size ($d \sim 0.3$) seems to suggest that within the Framework-oriented and Vague-oriented groups, students who perceived relevance scored slightly better than students who perceived irrelevance. This may offer a bit more detail to the Ref. 10 finding that achievement goals did not appear to correlate to FCI pre-post gains.

TABLE 2. Average FCI pretest and posttest score percentages and normalized gains, comparing between students with a "Relevant" view of physics and students with an "Irrelevant" view. All comparisons had equal variance (*F*-test: $p > 0.05$).

| FCI result | Irrelevant ($n = 50$) | Relevant ($n = 168$) | Cohen's *d* (*p*-value) |
|---|---|---|---|
| Pretest | 24.5% | 26.2% | 0.12 (0.44) |
| Posttest | 36.0% | 37.8% | 0.11 (0.49) |
| Gain (g) | +14.5 | +15.5 | 0.06 (0.70) |

### C. Effect of Relevance on CLASS Performance

As with the learning orientations in general (as discussed in Ref. 10), there is a likelihood that perceived relevance may affect students' pre-post attitudinal shifts. In Table 3, an investigation between "relevance" and "irrelevance" groups is examined for the CLASS survey, both overall and within each item cluster as defined by Adams et al.: [20] PI (Personal Interest), RWC (Real-World Connections), PS-G (Problem Solving – General), PS-C (Problem Solving – Confidence), PS-S (Problem Solving – Sophistication), SME (Sense-Making/Effort), CU (Conceptual Understanding), and ACU (Applied Conceptual Understanding). Comparsions featured are t-tests for statistical significance and Cohen's d values for effective size; all comparisons featured were determined to have statistically equal variances using F-tests.

The results of Table 3 seem to indicate a rather clear change from pretest to posttest. While perceived relevance does not show a statistical significance on the CLASS pretests, there is a slight trend of the students who perceive relevance having a slightly more expert-like attitude on the pretest. This becomes more statistically clear on the post-test, where students who perceive relevance are demonstrably more expert-like on almost all item clusters, and on average normalized gains, which clarify the uncertainty from the pre-tests by showing which sets of gains are statistically significant. There is a large effect size for post-test scores (d = 0.53) and a moderate effect size for normalized gains (d = 0.37) on the overall CLASS survey results. Only a few item clusters showed statistical significance for post-tests and normalized gains, namely Personal Interest, Real-World Connections, and Conceptual Understanding; the difference in normalized gains is moderate in all three cases (0.3 < d < 0.4), but the Personal Interest cluster shows a large effect size on post-test scores (d = 0.53).

The results of Table 3 appear to suggest that certain CLASS item clusters map in particular to students' perceived relevance of physics to their non-physical-science majors. In particular, the strong effect size for "Personal Interest" suggests that a perception that physics is irrelevant to one's major may be strongly linked to apathy towards physics. This also goes for "Real-World Connections," in that physics is not readily on a student's mind with regard to their major or to anything else in their worldview. Similarly, the achievement-goal-related learning orientations featured here were found to correspond to more expert-like shifts for Framework-oriented students than for Performance-oriented students and Vague-oriented students on the CLASS, overall and in particular for the Problem Solving item clusters. [10]

As discussed in Mason and Bertram, [10] biology majors appeared overall to be more prepared for the course and have more expert-like attitudes about learning the course; perhaps perception of relevance may vary between these two groups as well. Table 4 examines overall CLASS results within choice of life science major, in order to determine whether biology majors and health science majors may show different trends according to perceived irrelevance.

TABLE 3. Average CLASS pre-test and post-test percentages of expert-like scores and normalized gains, comparing between students who expressed "Relevance" and students who expressed "Irrelevance" in response to the post-test survey question "Which sections of the [class] material did you find related well to the coursework in your major?" The abbreviations for the item clusters are explained in Ref. 13. A positive gain indicates a net expert-like shift, and a negative gain indicates a net novice-like shift.

| Pretest scores ($n$) | Overall | PI | RWC | PS-G | PS-C | PS-S | SME | CU | ACU |
|---|---|---|---|---|---|---|---|---|---|
| Irrelevance % (50) | 54.3 | 44.0 | 64.0 | 58.8 | 58.5 | 39.7 | 65.7 | 51.0 | 37.4 |
| Relevance % (168) | 57.6 | 51.6 | 65.5 | 63.9 | 65.6 | 43.7 | 70.5 | 54.9 | 43.1 |
| Effect size | 0.22 | *0.29* | 0.05 | 0.23 | 0.26 | 0.15 | 0.22 | 0.15 | 0.25 |
| *p*-value | 0.18 | *0.08* | 0.76 | 0.16 | 0.11 | 0.34 | 0.18 | 0.36 | 0.13 |
| **Posttest scores ($n$)** | **Overall** | **PI** | **RWC** | **PS-G** | **PS-C** | **PS-S** | **SME** | **CU** | **ACU** |
| Irrelevance % (50) | 50.0 | 32.7 | 50.0 | 51.5 | 51.0 | 32.7 | 60.0 | 47.0 | 36.0 |
| Relevance % (168) | 58.1 | 47.8 | 63.2 | 61.5 | 62.9 | 42.5 | 68.1 | 56.5 | 43.5 |
| Effect size | **0.53** | **0.52** | *0.40* | *0.41* | *0.39* | *0.37* | *0.33* | *0.36* | *0.31* |
| *p*-value | **<0.01** | **<0.01** | **0.01** | **0.01** | **0.02** | **0.02** | **0.04** | **0.03** | *0.05* |
| **Normalized gains ($n$)** | **Overall** | **PI** | **RWC** | **PS-G** | **PS-C** | **PS-S** | **SME** | **CU** | **ACU** |
| Irrelevance % (50) | -0.06 | -0.24 | -0.15 | -0.04 | -0.03 | -0.21 | -0.03 | -0.10 | -0.14 |
| Relevance % (168) | +0.04 | -0.08 | +0.06 | +0.03 | +0.02 | -0.07 | +0.08 | +0.06 | -0.04 |
| Effect size | **0.37** | **0.33** | **0.36** | 0.14 | 0.10 | *0.28* | 0.24 | **0.33** | 0.22 |
| *p*-value | **0.02** | **0.04** | **0.03** | 0.38 | 0.56 | *0.08* | 0.14 | **0.04** | 0.18 |

TABLE 4. Average CLASS pretest and posttest percentages of expert-like scores and normalized gains, comparing students who perceived relevance vs. students who did not.

|  | Group | Irrelevant (*n*) | Relevant (*n*) |
|---|---|---|---|
| Pretest (%) | Biology | 55.7 (16) | 61.1 (75) |
|  | Health | 48.1 (19) | 52.3 (66) |
| Posttest (%) | Biology | **49.0 (16)** | **61.7 (75)** |
|  | Health | **43.4 (19)** | **52.5 (66)** |
| Gain (*g*) | Biology | **-.11 (16)** | **+.06 (75)** |
|  | Health | -.09 (19) | +.02 (66) |
| Cohen's *d* effect sizes (*p*-values from *t*-tests) | | | |
|  | **Biology** | | **Health** |
| Pretest | 0.39 (0.16) | | 0.29 (0.27) |
| Posttest | **0.79 (< 0.01)** | | **0.52 (< 0.05)** |
| Gains | **0.58 (< 0.05)** | | 0.39 (0.14) |

TABLE 5. Average CLASS pretest and posttest percentages of expert-like scores and normalized gains, comparing students who perceived relevance vs. students who did not.

|  | Group | Irrelevant (*n*) | Relevant (*n*) |
|---|---|---|---|
| Pretest (%) | Framework | 53.4 (17) | 58.1 (59) |
|  | Performance | 57.4 (16) | 57.1 (63) |
|  | Vague | 52.1 (17) | 57.6 (46) |
| Posttest (%) | Framework | *54.9 (17)* | *62.6 (59)* |
|  | Performance | *46.5 (16)* | *56.1 (63)* |
|  | Vague | 48.2 (17) | 54.9 (46) |
| Gains (*g*) | Framework | +.05 (17) | +.13 (59) |
|  | Performance | **-.18 (16)** | **+.01 (63)** |
|  | Vague | -.06 (17) | -.02 (46) |
| Cohen's *d* effect sizes (*p*-values from *t*-tests) | | | |
|  | **Framework** | **Performance** | **Vague** |
| Pretest | 0.35 (0.21) | 0.02 (0.94) | 0.35 (0.22) |
| Posttest | *0.47 (0.10)* | *0.53 (0.07)* | 0.40 (0.17) |
| Gains | 0.26 (0.35) | **0.68 (0.02)** | 0.18 (0.54) |

As Table 4 shows, however, the statistical significance appears for both biology majors and health science majors on the CLASS post-test. As for normalized gains, biology majors who perceive relevance show statistically more expert-like gains than students who do not, while health science majors do not demonstrate this statistical significance and only show a moderate effect size.

Table 4 indicates that each life science major separately exhibits the pattern within the overall CLASS results in Table 3, between relevant-perceiving and irrelevant-perceiving students. Both sets of majors show no statistical significance on the pretest, but a large effect size and significance on the posttest (with more expert-like scores from relevant-perceiving students in both cases). With regard to CLASS item clusters, biology majors who perceived relevance had statistically higher post-test scores and normalized gains on the PI and RWC clusters, as well as on the Sense Making-Effort (SME) cluster, with large effect sizes for these three clusters ($0.6 < d < 0.85$ for post-test comparisons, $0.5 < d < 0.8$ for gains comparisons).

Table 5 shows CLASS results within learning orientations between perceived relevance and irrelevance. We are particularly interested to see if learning orientations and perceived relevance influence each other, as it appears both items can be considered factors that influence attitudinal shifts. However, perceived relevance appears to only be a factor for Performance-oriented students; those who perceived relevance to their majors did not experience much shifting, while those who did not experienced a novice-like shift. The effect size between relevance and irrelevance for Performance-oriented students is large (d = 0.68). On the other hand, Framework-oriented students are only mildly influenced by perceived relevance (d = 0.26 on gains), with a slightly expert-like shift occurring for both relevance and irrelevance groups; Vague-oriented students also had a slight novice-like shift regardless of perceived relevance. Within the CLASS item clusters, the Performance-oriented students had statistical significance ($p < 0.05$) or borderline significance ($0.05 < p < 0.1$) across all item clusters, with students who perceived irrelevance having a more novice-like shift in each case.

## V. DISCUSSION

### A. Conclusions

In terms of the first research question, namely whether students' perceived relevance had an effect on their FCI pre-post outcomes, the null argument that there is no difference on the FCI appears to be accepted, at least within the overall sample size of 218 students as split into 168 students who perceived at least some form of relevance and 50 students who explicitly said they did not. There was no statistical significance between groups on either the pretest or the posttest, and the effect sizes on both the pretest and the protest were weak. A subdivision into item clusters, choice of life science major, or learning goals, as was done for CLASS results, did not show any significant results either, and so the data from that particular comparison is omitted.

As Table 2's normalized gains seem to strongly indicate, the larger factors appear to be the particularly low pretest scores from this sample (which are close to 20%, amounting to the average score for randomly choosing from 5 answers for each of the multiple-choice questions). This in turn skews the normalized gain calculation, as the denominator of this calculation (100% - pre score) is larger than are the pretest scores of many calculus-based introductory physics courses' gains, whether within traditional instruction or within research-based course transformations. [38]

With regard to the second research question, on the other hand, there is a strong difference on CLASS pre-post attitudinal shifts. While students who perceived relevance were statistically similar to students who did not perceive relevance at the beginning of the course, the pre-post shifts (shown both by post-test results and by normalized gains) demonstrate that students who do not perceive relevance experience a slight novice-like shift overall, as opposed to a slight expert-like shift for students who do perceive relevance. In particular, the strongest results within the CLASS item clusters appear within the Personal Interest and Real-World Connections clusters. This appears to corroborate students' self-reported perceptions that nothing in the physics course applies to their major, by suggesting that this group of students is a) less likely to be interested in physics (a lack of perceived relevance may be conflated with a more straightforward lack of interest in the subject) and b) less likely to perceive real-world connections (including to their own majors) of physics.

We move from here into the third research question, namely whether perceived relevance has an effect within choice of major or within learning orientation. An analysis of relevance vs. irrelevance within both biology majors and health science majors appears to show a slightly stronger trend for biology majors than for health science majors. While both major groups demonstrate a statistical difference in posttest results, only the biology majors show statistical difference on normalized gains, while health science majors do not. That having been said, perceived relevance does appear to matter for both sets of life science majors.

As for achievement-goal-based learning orientations, perceived relevance appears to make a difference specifically for performance-oriented students. Those that perceive relevance of physics to their majors do not experience a novice-like shift (or for that matter, much of a shift at all), while those that do not perceive relevance experience a more pronounced novice-like shift. Therefore, even though some students may be more concerned about their course grade than about mastering the material, they may at least retain their attitudes towards learning physics without a novice-like shift if they can perceive that at least something within a physics class pertains to their major. On the other hand, perceived relevance only seems to have a mild relative effect within Framework-oriented students (whose focus on mastery gives them an expert-like shift in either case of perceiving relevance or not) and within Vague-oriented students (who do not state any achievement goal and who experience a novice-like shift in either case).

Overall, perception of relevance appears to be a co-factor alongside achievement-goals-driven learning orientations for the expert-novice shifts within attitudinal survey data. It appears that aspects of student mindset do influence perceptions of learning physics; however, it is not so clear whether student mindset affects conceptual understanding (at least with regard to the force and motion topics within the FCI).

### B. Limitations of the Study

The study's results are limited to the respective scopes of the two survey instruments, namely the FCI and the CLASS. The FCI focuses on force and motion topics, not necessarily upon energy topics or other topics that may be covered within a first-semester course. The low pretest scores on the FCI also suggest that the student population at the host institution (a regional four-year state university with a primary emphasis on undergraduate education) may not match student populations in published studies at more research-oriented institutions.

There is also the limitation of the host university's status as a primarily undergraduate institution with a regional focus. This presented logistic issues that prevented collection of a large sample size, namely that only 48-72 students were available to the author for any given semester, such that multiple semesters were required to collect at least 200 students. In addition, the institution type influences both the meaning of the results (in that the students involved do not necessarily have the same demographics, career goals, etc., as e.g. students in a large research university) and the feasibility of recommended means to address the problem. For an example of the latter issue, the featured health science majors were a mixture of majors that somewhat resembled pre-medical and other human-anatomy-related biology tracks (e.g. pre-physical therapy and pre-occupational therapy), while other majors did not even clearly have a physical science application at all (e.g. the health administration track is akin to a business major). With these allowances in mind, the results may serve more useful for non-research-oriented four-year state institutions than for other institution types.

The post-test feedback survey had its own limitations as well. First, as it was not a validated research instrument, the findings with the CLASS survey's Personal Interest and Real-World Connections item clusters serve as a form of validation, in the sense that perceived relevance is closely related to the expert-novice shifts within these two clusters. Second, because the feedback survey was not a pre-test instrument, and because there were no statistically significant differences between groups on the CLASS in either case, it is unknown whether either perceived relevance or achievement goals are static or growth-related throughout the semester. [1,9] The relative trends on the CLASS pre-test appear to suggest that, with a larger sample size, it could be roughly determined whether perceived relevance may be more likely to be static or growth-related.

## C. Discussion

Even given the limitations within the FCI data, the results to reinforce the notion that expert-like beliefs are not necessarily correlated with strong content understanding of physics. The CLASS is a validated instrument for students' perceptions of learning physics; that being said, there have been recent discussions about whether attitudinal changes actually reflect students' ability to perform well in the course, e.g. item response theory explorations of the CLASS survey which suggest that high performance in the course does not necessarily correlate with expert-like beliefs. [39] Previous work that looked more closely at this student sample's overall course grades [10] showed only a borderline significant argument that students who stated an achievement goal (either mastery-related or performance-related) were somewhat more successful than students who did not (i.e. the Vague-oriented group). Due to the IRB limitations on this follow-up study, the author cannot provide a similar discussion of course grades' relationship with perceived course relevance; a follow-up study with newer data and the ability to measure this will shed light on this issue. If it turns out that perceived relevance is related to course performance, the next goal would be to investigate how to convince students over the course of the semester that introductory physics is indeed relevant to their majors; if it does not, then the issue may simply be that a single introductory-level course is not a large enough scope to observe the effects of perceived relevance on course grades, and a more longitudinal study may be necessary.

## ACKNOWLEDGEMENTS

The author thanks the UCA Department of Physics and Astronomy for funding support.